\documentstyle[12pt]{article}
\begin{document}
\title{Pair production and solutions of the wave equation for particles
with arbitrary spin\thanks{Invited talk at the international workshop
"Lorentz Group, CPT, and Neutrinos", Zacatecas, Mexico, June 23-26, 1999.}}
\author{S.I. Kruglov\thanks{E\_mail: skruglov23@hotmail.com}
\\ \it International Education Centre, 2727 Steeles Ave. W, \# 202,
\\ \it Toronto, ON M3J 3G9, Canada
\\ \it
On leave from
\\ \it National Scientific Center of
Particle and High Energy Physics,
\\ \it
M. Bogdanovich St. 153, Minsk 220010, Belarus}
\maketitle

\begin{abstract}
We investigate the theory of particles with arbitrary spin and anomalous
magnetic moment in the Lorentz representation $\left( 0,s\right)
\oplus \left( s,0\right)$, in an external constant and uniform electromagnetic
field. We obtain the density matrix of free particles in pure spin
states. The differential probability of pair producing particles with arbitrary
spin by an external constant uniform electromagnetic field has been found
using the exact solutions. We have calculated the imaginary and real parts
of the Lagrangian in an electromagnetic field that takes into account
vacuum polarization.
\end{abstract}

\section{Introduction}

The interest to the theory of relativistic particles with arbitrary spins is
growing now. One of the reason is that SUSY models require superpartners,
i.e. additional fields of particles with higher spins. In particular, it is
important to take into account particles with spin $3/2$ - gravitons for the
theory of inflation of the universe [1]. The second aspect is that some of
string models have the similar features as models of relativistic spinning
particles [2]. It is also interesting to have particles with arbitrary
fractional spins [3] (see also [4, 5]). Such spinning particles in $2+1$
dimensions, called anyons, were discovered and they have anomalous
statistics.

There are many different relativistic wave equations which describe
particles with arbitrary spin [6-11]. The fields of free particles of a
mass $m$ and spin $s$ in these formalisms realize definite representations
of the Poincar\'e group. Some of these schemes for particles are equivalent
each other. If the interaction with external electromagnetic fields is
introduced, approaches based on different representations of the Lorentz
group become not equivalent. Most of theories of particles in the external
electromagnetic fields have difficulties such as non-causally propagation
[12], indefinite metrics in the second-quantized theory [13-15] and others.
We proceed here from the second order of the relativistic wave equation for
particles with arbitrary spin $s$ and magnetic moment $\mu $ on the base of
the Lorentz representation $(0,s)\oplus (s,0)$, where $\oplus $ means the
direct sum. Such an approach avoids some difficulties as compared with
others schemes and the corresponding wave function has the minimal number of
components. This is a generalization of the Feynman-Gell-Mann equation [16]
for particles with spin $1/2$ in the case of higher spin particles which
possess arbitrary magnetic moment. In particular case of spin $1/2$ we
arrive to the Dirac theory. If the normal magnetic moment is considered, it
leads to the approach [11]. Particles in this scheme propagate causally in
external electromagnetic fields and this is a parity-symmetric theory with a
Lagrangian formulation. The authors of the works [17] emphasized the
importance of approaches based on the $(0,s)\oplus (s,0)$ representation of
the Lorentz group in the light of new experimental data. In [11] $(6s+1)$
-component first-order matrix formulation of the equation for particles with
arbitrary spin was considered. Then the author received the second order
equation for particles with ''normal'' magnetic moment. Starting with the
second order equation for particles which possess arbitrary magnetic moment
we transit to the first order wave equation with another representation of
the Lorentz group. But algebraic properties of the matrices are the same as
in the approach [11]. We found solutions of equations for free particles in
the form of density-matrices (projective matrices-dyads) for pure spin
states which are used for different electromagnetic calculations of the
Feynman diagrams. Such projective matrices-dyads allow to make covariant
calculations without using matrices of the wave equation in the definite
representation.

The main purpose of this paper is to investigate solutions of wave
equations, pair production of arbitrary spin particles by constant uniform
electromagnetic fields and vacuum polarization of higher spin particles.
Considering one-particle theory and obtaining the differential probability
for pair production of particles with arbitrary spins we avoid the Klein
paradox [18, 19]. As particular case of spin $1/2$ and gyromagnetic ratio $2$
particles we arrive at the well-known result found by Schwinger [22] who
predicted $e^{+}e^{-}$ pair production in the strong external
electromagnetic field. It is actually now in the light of the development of
power laser techniques. It should be noted that the pair production of
particles by a gravitational field also is important for understanding the
evolution of the universe near singularity [20].

The probability of pair production of particles in external electromagnetic
fields can be found on the base of exact solutions of the wave equations
[21] or the imaginary part of the Lagrangian [22]. We consider here both
approaches. Nonlinear corrections to the Maxwell Lagrangian of the constant
uniform electromagnetic fields are determined from the polarization of the
vacuum of arbitrary spin particles. The problem of pair production of
particles with higher spins using the quasiclassical scheme (method of
''imaginary time'') was considered in [23] which is agreed with our approach
of the relativistic wave equations. It should be noted that the
quasiclassical approximation has a restriction for the fields considered $E,$
$H\ll m^2/e$ when the process is $\exp $onentialy suppressed. It means that
the approach [23] is valid when electromagnetic fields are not too strong
and less than the critical value $m^2/e$. But it is known that pair of
particles are created rapidly at the critical value of the fields. In our
consideration there are no such restrictions. The problem of the pair
production of vector particles with gyromagnetic ratio $2$ was investigated
in [24]. In [25, 26] the imaginary part of the effective Lagrangian which
defines the probability of $e^{+}e^{-}$-production was found with taking
into account anomalous magnetic moment.

We use system of units $\hbar =c=1$, $\alpha =e^2/4\pi =1/137$, $e>0$. In $
\sec $tion 2 proceeding from the second order equation for arbitrary spin
particles with anomalous magnetic moment we transfer to the first order
formulation of the theory. All independent solutions of the equation for
free particles are found in the form of matrices-dyads (density matrix).
Section 3 contains investigating of exact solutions of arbitrary spin
particle equations in the constant uniform electromagnetic fields. We find
on the base of exact solutions the differential probability for pair
production of particles with the arbitrary spin and anomalous magnetic
moment. The imaginary part of the effective Lagrangian for the
electromagnetic fields is calculated. In section 4 using the Schwinger
method we find the nonlinear corrections to the Lagrangian of a constant
uniform electromagnetic fields, caused by the vacuum polarization of
particles with the arbitrary spin and magnetic moment. Section 5 contains
the discussion of results.

\section{ Wave Equation and Density Matrix}

We proceed here from the Bargmann-Wightman-Wigner-type (BWW-type) quantum
field theory [27, 17] based on the $(s,0)\oplus (0,s)$ Lorentz
representation for massive particles. In the BWW-type theories, a boson and
antiboson have opposite intrinsic parities [28] and well-defined $C$ and $T$
characteristics [17]. The wave function of the $(s,0)\oplus (0,s)$
representation has $2(2s+1)$ component. For the spin $1/2$ we arrive at
well-known Dirac bispinors. But for spin $1$ there is doubling of the
component as compared with the Proca theory [29] because the vector
particles have $3$ spin states with the projections $s_z=\pm 1,0$. As
pointed in [17] the $(s,0)\oplus (0,s)$ representation for massive particles
opens new experimentally observable possibilities.

We postulate the next two (for $\varepsilon =\pm 1$) wave equations for
arbitrary spin particles in external electromagnetic fields:
\begin{equation}
(D_\mu ^2-m^2-\frac{eq}{2s}F_{\mu \nu }\Sigma _{\mu \nu }^{(\varepsilon
)})\Psi _\varepsilon (x)=0,
\end{equation}
where $s$ is the spin of particles, $D_\mu =\partial _\mu -ieA_\mu$ ;$F_{\mu
\nu }=\partial _\mu A_\nu -\partial _{_\nu }A_\mu $ is the strength tensor, $
\varepsilon =\pm 1;$ and $\Sigma _{\mu \nu }^{\left( -\right) }$, $\Sigma
_{\mu \nu }^{\left( +\right) }$ are the generators of the Lorenz group which
correspond to the $\left( s,0\right) $ and $\left( 0,s\right)$
representations. Two equations (1) (for $\varepsilon =\pm 1$) describe
particles which possess the magnetic moment $\mu =eq/(2m)$ and gyromagnetic
ratio $g=q/s$. At $q=1$ we have ''normal'' magnetic moment $\mu =e/(2m)$ and 
$g=1/s$. Generators $\Sigma _{\mu \nu }^{(\varepsilon )}$ are connected with
the spin matrices $S_k$ by the relationships $\Sigma _{ij}^{(\varepsilon
)}=\varepsilon _{ijk}S_k$, $\Sigma _{4k}^{(\varepsilon )}=-i\varepsilon S_{k}
$, where the parameter $\varepsilon =\pm 1$ corresponds to the Lorentz group
representations $\left( s,0\right) $ and $\left( 0,s\right) $. As usual,
relations
\begin{equation}
\left[ S_i,S_j\right] =i\varepsilon _{ijk}S_k,\hspace{1.0in}\left(
S_1\right) ^2+\left( S_2\right) ^2+\left( S_3\right) ^2=s(s+1)
\end{equation}
are valid, where $i,j,k=1,2,3;\varepsilon _{ijk}$ is the Levi-Civita symbol.

At $q=1$ equations (1) where considered in [11]. The theory of arbitrary
spin particles based on equations (1) is causal in the presence of external
electromagnetic fields. It is seen from the method [12] that equations
remain hyperbolic and the characteristic surfaces are lightlike. Equations
(1) are invariant at the parity operation. Indeed, at the parity inversion $
\varepsilon \rightarrow -\varepsilon $ and the representation $(s,0)$
transforms into $(0,s)$.

Now we consider the problem to formulate a first-order relativistic wave
equation from the second order equation (1). This is convenient for some
quantum electrodynamics calculations with polarized particles of arbitrary
spins.

Let us introduce the matrix $\varepsilon ^{A,B}$ which has the dimension $
n\times n$, its elements consists of zeroes and only one element is unit on
the crossing of row $A$ and column $B$. So the matrix element and
multiplication of this matrices are
\begin{equation}
\left( \varepsilon ^{A,B}\right) _{CD}=\delta _{AC}\delta _{BD}, 
\hspace{1.0in}\varepsilon ^{A,B}\varepsilon ^{C,D}=\delta _{BC}\varepsilon
^{A,D},
\end{equation}
where indices $A,B,C,D=1,2...n$.

The six generators $\Sigma _{\mu \nu }^{(+)}$ (or $\Sigma _{\mu \nu }^{(-)}$)
entering equation (1) have the dimension \mbox{$2s+1$}. Therefore the wave function 
$\Psi _{+}(x)$ (or $\Psi _{-}(x)$) of equation (1) possesses $2s+1$
components. Now we can introduce the $5(2s+1)$-component wave function
\begin{equation}
\Psi _1(x)=\left( 
\begin{array}{c}
\Psi _{+}(x) \\ 
-\frac 1mD_\mu \Psi _{+}(x)
\end{array}
\right)
\end{equation}
so that $\Psi _1(x)=\left\{ \Psi _A(x)\right\} ,$ $A=0,\mu ;$ $\Psi _0=\Psi
_{+}(x),$ $\Psi _\mu =-\frac 1mD_\mu \Psi _{+}(x)$ and $\Psi _{+}(x)$
realizes the Lorentz representation $(s,0)$. It is not difficult to check
that equation (1) for $\varepsilon =+1$ can be represented as the
first-order equation
\begin{equation}
\left( \beta _\mu ^{(+)}D_\mu +m\right) \Psi _1(x)=0,
\end{equation}
where $5(2s+1)\times 5(2s+1)$- matrices
\begin{equation}
\beta _\mu ^{(+)}=\left( \varepsilon ^{0,\mu }+\varepsilon ^{\mu ,0}\right)
\otimes I_{2s+1}-i\frac qs\varepsilon ^{0,\nu }\otimes \Sigma _{\mu \nu
}^{(+)},
\end{equation}
are introduced and $\otimes $ is the direct multiplication, $I_{2s+1}$ is
the unit matrix of the dimension $2s+1$ and in (6) we imply the summation on
index $\nu $. Using properties (3) it is easy to check that $5$-dimensional
matrices $\beta _\mu ^{DK}=\varepsilon ^{0,\mu }+\varepsilon ^{\mu ,0}$ obey
the Duffin-Kemmer algebra [30, 31]
\begin{equation}
\beta _\mu ^{DK}\beta _\nu ^{DK}\beta _\alpha ^{DK}+\beta _\alpha ^{DK}\beta
_\nu ^{DK}\beta _\mu ^{DK}=\delta _{\mu \nu }\beta _\alpha ^{DK}+\delta
_{\alpha \nu }\beta _\mu ^{DK}.
\end{equation}

The wave function $\Psi _1(x)$ transforms on the $\left[ \left( 0,0\right)
\oplus \left( 1/2,1/2\right) \right] \otimes (s,0)$ representation of the
Lorentz group [32,33]. For the case $\varepsilon =-1$ we have the analogous
equation
\begin{equation}
\left( \beta _\mu ^{(-)}D_\mu +m\right) \Psi _2(x)=0,
\end{equation}
where 
\[
\beta _\mu ^{(-)}=\left( \varepsilon ^{\overline{0},\overline{\mu }
}+\varepsilon ^{\overline{\mu },\overline{0}}\right) \otimes I_{2s+1}-i\frac
qs\varepsilon ^{\overline{0},\overline{\nu }}\otimes \Sigma _{\mu \nu
}^{(-)}, 
\]
\begin{equation}
\Psi _2(x)=\left( 
\begin{array}{c}
\Psi _{-}(x) \\ 
-\frac 1mD_\mu \Psi _{-}(x)
\end{array}
\right)
\end{equation}
and $\Psi _2(x)=\left\{ \Psi _B(x)\right\} $, $B=\overline{0},\overline{\mu};
$ $\Psi _{\overline{0}}(x)=\Psi _{-}(x)$, $\Psi _{\overline{\mu }}=-\frac
1mD_\mu \Psi _{-}(x)$ where $\Psi _{-}(x)$ transforms as $(0,s)$
representation of the Lorentz group and $\Psi _2(x)$ realizes the representation
\mbox{$\left[
\left( 0,0\right) \oplus \left( 1/2,1/2\right) \right] \otimes (0,s)$}.
The two equations (5) and (7) can be combined into one first-order
equation
\begin{equation}
\left( \beta _\mu ^{}D_\mu +m\right) \Psi (x)=0
\end{equation}
with the matrices and wave function
\begin{equation}
\beta _\mu ^{}=\beta _\mu ^{(+)}\oplus \beta _\mu ^{(-)},\hspace{1.0in}\Psi
(x)=\left( 
\begin{array}{c}
\Psi _1(x) \\ 
\Psi _2(x)
\end{array}
\right) .
\end{equation}

Using the properties of elements of the entire algebra (3), it is not
difficult to get the relation equation for matrices $\beta _\mu ^{}$ (the
same for $\beta _\mu ^{(+)}$ and $\beta _\mu ^{(-)}$): 
\[
\beta _\mu ^{}\beta _\nu ^{}\beta _\sigma ^{}+\beta _\nu ^{}\beta _\sigma
^{}\beta _\mu ^{}+\beta _\sigma ^{}\beta _\mu ^{}\beta _\nu ^{}+\beta _\nu
^{}\beta _\mu ^{}\beta _\sigma ^{}+\beta _\mu ^{}\beta _\sigma ^{}\beta _\nu
^{}+\beta _\sigma ^{}\beta _\nu ^{}\beta _\mu ^{}= 
\]
\begin{equation}
=2\left( \delta _{\nu \sigma }\beta _\mu ^{}+\delta _{\mu \sigma }\beta _\nu
^{}+\delta _{\mu \nu }\beta _\sigma ^{}\right) .
\end{equation}

In works [11] the $(6s+1)$-dimensional representation of $SL(2,C)$ group was
considered for particles of arbitrary spins with algebra (12). It is seen
from (12) that the algebra of $\beta _\mu ^{}$ matrices is more complicated
then Duffin-Kemmer algebra (7). Different representations of this algebra
were studied in [34, 35].

Let us consider the problem to find the solutions to equation (10) for the
definite momentum and spin projections. It is convenient to find these
solutions in the form of projective matrix-dyads (density matrix). All
electrodynamics calculations of Feynman diagrams with arbitrary spin
particles can be done using these matrices. As particles in initial and
final states are free particles, we can put parameter $q=0$ in (1), (10). It
corresponds to the case when external electromagnetic fields are absent.
Than matrices $\beta _\mu ^{}$ transforms to $\beta _\mu ^0:$

\begin{equation}
\beta _\mu ^0=\left[ \left( \varepsilon ^{0,\mu }+\varepsilon ^{\mu
,0}\right) \otimes I_{2s+1}\right] \oplus \left[ \left( \varepsilon ^{
\overline{0},\overline{\mu }}+\varepsilon ^{\overline{\mu },\overline{0}
}\otimes I_{2s+1}\right) \right] ,
\end{equation}
which obey the Duffin-Kemmer algebra (7). The projective operators
extracting states with definite 4-momentum $p_\mu $ for particle and
antiparticle are given by
\begin{equation}
\Lambda _{\pm }=\frac{i\widehat{p}\left( i\widehat{p}\pm m\right) }{2m},
\end{equation}
where $\widehat{p}=p_\mu \beta _\mu ^0$ (we use the metric such that $p^2= 
{\bf p}^2+p_4^2={\bf p}^2-p_0^2=-m^2$). Signs $+$ and $-$ in (14) correspond
to the particle and antiparticle, respectively. Matrices $\Lambda _{\pm }$
have the usual projective operator properties [36]
\begin{equation}
\Lambda _{\pm }^2=\Lambda _{\pm }.
\end{equation}

Equation (15) is checked by the relation $\widehat{p}^3=p^2\widehat{p}$
which follows from the Duffin-Kemmer algebra (7). To find the spin
projective operators we need the generators of the Lorentz group in the
representation of the wave function $\Psi (x)$ which enters equation~(10).
From the structure of the functions $\Psi _1(x)$, $\Psi _2(x)$ (4), (9) we
define the generators of the Lorentz group in our $10(2s+1)$-dimension
representation 
\[
J_{\mu \nu }=J_{\mu \nu }^{(+)}\oplus J_{\mu \nu }^{(-)}, 
\]
\begin{equation}
J_{\mu \nu }^{(+)}=\left( \varepsilon ^{\mu ,\nu }-\varepsilon ^{\nu ,\mu
}\right) \otimes I_{2s+1}+iI_5\otimes \Sigma _{\mu \nu }^{(+)},
\end{equation}
\[
J_{\mu \nu }^{(+)}=\left( \varepsilon ^{\overline{\mu },\overline{\nu }
}-\varepsilon ^{\overline{\nu },\overline{\mu }}\right) \otimes
I_{2s+1}+iI_5\otimes \Sigma _{\mu \nu }^{(-)}, 
\]
where $I_5$ is the $5$-dimensional unit matrix. Using properties (3), we get
the commutation relations

\begin{equation}
\left[ J_{\mu \nu }^{},J_{\alpha \beta }^{}\right] =\delta _{\nu \alpha
}J_{\mu \beta }^{}+\delta _{\mu \beta }J_{\nu \alpha }^{}-\delta _{\mu
\alpha }J_{\nu \beta }^{}-\delta _{\nu \beta }J_{\mu \alpha }^{},
\end{equation}
\begin{equation}
\left[ \beta _\mu ^{},J_{\alpha \beta }^{}\right] =\delta _{\mu \alpha
}\beta _\beta ^{}-\delta _{\mu \beta }\beta _\alpha ^{}.
\end{equation}

The elationship (17) is a well-known commutation relation for generators of the
Lorentz group [32, 33]. Equation (10) is a form invariant under the Lorentz
transformations because relation (18) is valid. To guarantee the existence
of a relativistically invariant bilinear form

\begin{equation}
\overline{\Psi }\Psi =\Psi ^{+}\eta \Psi ,
\end{equation}
where $\Psi ^{+}$ is the Hermite conjugated wave function, we should
construct a Hermitianizing matrix $\eta $ with the properties [34, 36]:

\begin{equation}
\eta \beta _i=-\beta _i\eta ,\hspace{1.0in}\eta \beta _4=\beta _4\eta 
\hspace{1.0in}(i=1,2,3).
\end{equation}

Such matrix exists and is given by
\begin{equation}
\eta =\left( \varepsilon ^{a,\overline{a}}+\varepsilon ^{\overline{a}
,a}-\varepsilon ^{4,\overline{4}}-\varepsilon ^{\overline{4},4}-\varepsilon
^{0,\overline{0}}-\varepsilon ^{\overline{0},0}\right) \otimes I_{2s+1},
\end{equation}
where the summation on the index $a=1,2,3$ is implied. Now we introduce the
operator of the spin projection on the direction of the momentum ${\bf p}$ :

\begin{equation}
\widehat{S}_{{\bf p}}=-\frac i{2\mid {\bf p\mid }}\varepsilon
_{abc}p_aJ_{bc}=\left( \kappa _p+\sigma _p\right) \oplus \left( \overline{
\kappa }_p+\overline{\sigma }_p\right) ,
\end{equation}
where 
\[
\kappa _p=-\frac i{\mid {\bf p\mid }}\varepsilon _{abc}p_a\varepsilon
^{b,c}\otimes I_{2s+1},\hspace{1.0in}\overline{\kappa }_p=-\frac i{\mid {\bf 
p\mid }}\varepsilon _{abc}p_a\varepsilon ^{\overline{b},\overline{c}
}\otimes I_{2s+1}, 
\]

\begin{equation}
\sigma _p=\overline{\sigma }_p=I_5\otimes \frac{{\bf pS}}{\mid {\bf \ p\mid }
},
\end{equation}
and $\mid {\bf p\mid =}\sqrt{p_1^2+p_2^2+p_3^2}.$ It is easy to check that
the required relation holds 
\[
\left[ \widehat{S}_{{\bf p}},\widehat{p}\right] =0. 
\]

The matrices $\kappa _p$, $\overline{\kappa }_p$ obey the simple equations
\begin{equation}
\kappa _p^3=\kappa _p,\hspace{1.0in}\overline{\kappa }_p^3=\overline{\kappa }
_p.
\end{equation}
Taking into account equations (2) we write out relations for the matrices $
\sigma _p$ (23): 
\[
\left( \sigma _p^2-\frac 14\right) \cdot \cdot \cdot \left( \sigma
_p^2-s^2\right) =0\hspace{1.0in} for~ odd~ spins, 
\]

\begin{equation}
\sigma _p^{}\left( \sigma _p^2-1\right) \cdot \cdot \cdot \left( \sigma
_p^2-s^2\right) =0\hspace{1.0in} for~ even~ spins.
\end{equation}

Relations (25) allow us to construct projective operators which extract the
pure spin states. Using the relationship 
\[
\widehat{S}_{{\bf p}}\widehat{p}=\left( \sigma _p\oplus \overline{\sigma }
_p\right) \widehat{p} 
\]
we can consider projective matrices on the base of equations (25). The
common technique of the construction of such operators is described in [36].
Let us consider the equation for auxiliary spin operators $\sigma _p$, $
\overline{\sigma }_p$ for spin projection $s_k:$

\begin{equation}
\sigma _p^{}\Psi _k=s_k\Psi _k.
\end{equation}

The solution to equation (26) can be found using relationships (25) which
can be rewritten as
\begin{equation}
\left( \sigma _p^{}-s_k\right) P_k(s)=0,
\end{equation}
where polynomials $P_k(s)$ are given by

\[
P_k(s)=\left( \sigma _p^2-\frac 14\right) \cdot \cdot \cdot \left( \sigma
_p^{}+s_k\right) \cdot \cdot \cdot \left( \sigma _p^2-s^2\right) 
\hspace{0.3in}for~odd~spins,
\]
\begin{equation}
P_k(s)=\sigma _p^{}\left( \sigma _p^2-1\right) \cdot \cdot \cdot \left(
\sigma _p^{}+s_k\right) \cdot \cdot \cdot \left( \sigma _p^2-s^2\right) 
\hspace{0.3in}for~even~spins.
\end{equation}

Every columns of the polynomial $P_k(s)$ can be considered as eigenvector $
\Psi _k$ of equation (26) with the eigenvalue $s_k$. As $s_k$ is one
multiple root of equations (25), all columns of the matrix $P_k(s)$ are
linear independent solutions of equation (26) [36]. Using definitions (28)
it can be verified that matrix
\begin{equation}
Q_k=\frac{P_k(s)}{P_k(s_k)}
\end{equation}
is the projective operator with the relation
\begin{equation}
Q_k^2=Q_k^{}.
\end{equation}

Equation (30) tells that the matrix $Q_k$ can be transformed into diagonal
form, where on the diagonal there are only units and zeroes. So the $Q_k$
acting on the wave function $\Psi $ will remain components which correspond
to the spin projection $s_k.$

We have mentioned that this theory of arbitrary spin particles has doubling
of spin states of particles because there are two representations $(s,0)$
and $(0,s)$ of the Lorentz group. To differ this representations (for $s>1/2$)
which are connected by the parity transformations we use the parity operator

\begin{equation}
K=\left( \varepsilon ^{\mu ,\overline{\mu }}+\varepsilon ^{\overline{\mu }
,\mu }+\varepsilon ^{0,\overline{0}}+\varepsilon ^{\overline{0},0}\right)
\otimes I_{2s+1}
\end{equation}
with the summation on indices $\mu =1,2,3,4.$ The $10(2s+1)$ matrix $K$ has
the property $K^2=I_{10(2s+1)}.$ The projective operator extracting states
with the definite parity is given by

\begin{equation}
M_\varepsilon =\frac 12\left( K+\varepsilon \right) ,
\end{equation}
where $\varepsilon =\pm 1$. This matrix possesses the required relationship

\begin{equation}
M_\varepsilon ^2=M_\varepsilon .
\end{equation}

It should be noted that the matrix $K$ (31) plays the role analogous to the $
\gamma _5$-matrix in the Dirac theory of particles with the spin $1/2$. It
is checked that operators $\widehat{p},$ $\widehat{S}_{{\bf p}}$, $K$
commute each other and as a consequence, they have the common eigenvector.
The projective operator extracting pure state with the definite 4-momentum
projections, spins and parity is given by

\begin{equation}
\Pi _{\pm m,k,\varepsilon }=\Lambda _{\pm }M_\varepsilon \left( Q_k\oplus
Q_k\right) 
\end{equation}
with matrices (14), (29) and (32). The $\Pi _{\pm m,k,\varepsilon }$ is the
density matrix for pure states. It is easy to consider not pure states by
summating (34) over definite quantum numbers $s_k$, $\varepsilon .$
Projective operator for pure states can be represented as matrix-dyad [36]:

\begin{equation}
\Pi _{\pm m,k,\varepsilon }=\Psi _{\pm m,k}^\varepsilon \cdot \overline{\Psi 
}_{\pm m,k}^\varepsilon ,
\end{equation}
where $\overline{\Psi }_{\pm m,k}^\varepsilon =\left( \Psi _{\pm
m,k}^\varepsilon \right) ^{+}\eta $ and the $\Psi _{\pm m,k}^\varepsilon $
is the solution to equations 
\[
\left( i\beta _\mu ^{}p_\mu \pm m\right) \Psi _{\pm m,k}^\varepsilon =0, 
\hspace{1.0in}\widehat{S}_{{\bf p}}\Psi _{\pm m,k}^\varepsilon =s_k\Psi
_{\pm m,k}^\varepsilon , 
\]

\begin{equation}
K\Psi _{\pm m,k}^\varepsilon =\varepsilon \Psi _{\pm m,k}^\varepsilon .
\end{equation}

Expression (35) is convenient for calculations of different quantum
electrodynamics processes with polarized particles of arbitrary spins.

\section{Pair Production by External Electromagnetic Fields}

To calculate the probability of pair production of arbitrary spin particles,
we follow the Nikishov method [21]. So exact solutions to equation (1)
should be found for external constant uniform electromagnetic fields. Using
the properties of generators $\Sigma _{\mu \nu }^{(\varepsilon )}$ we find
the relationships

\begin{equation}
\frac 12\Sigma _{\mu \nu }^{(+)}F_{\mu \nu }=S_iX_i,\hspace{1.0in}\frac
12\Sigma _{\mu \nu }^{(-)}F_{\mu \nu }=S_iX_i^{*},
\end{equation}
where $X_i=H_i+iE_i$, $X_i^{*}=H_i-iE_i;$ $E_i,$ $H_i$ are the electric and
magnetic fields, respectively, and the spin matrices $S_i$ obey equations
(2). In the diagonal representation, the equations for eigenvalues are given
by
\begin{equation}
S_iX_i\Psi _{+}^{(\sigma )}(x)=\sigma X\Psi _{+}^{(\sigma )}(x), 
\hspace{1.0in}S_iX_i^{*}\Psi _{-}^{(\sigma )}(x)=\sigma X^{*}\Psi
_{-}^{(\sigma )}(x),
\end{equation}
where $X=\sqrt{{\bf X}^2},$ ${\bf X=H+}i{\bf E}$, and the spin projection $
\sigma $ is
\begin{equation}
\sigma = 
\begin{array}{c}
\pm s,\pm (s-1),\cdot \cdot \cdot 0 \\ 
\pm s,\pm (s-1),\cdot \cdot \cdot \pm \frac 12
\end{array}
\hspace{1.0in} 
\begin{array}{c}
for~ even~ spins, \\ 
for~ odd~ spins.
\end{array}
.
\end{equation}

Taking into account (37), (38), equations (1) (for $\varepsilon =\pm 1$) are
rewritten as
\begin{equation}
(D_\mu ^2-m^2-a\sigma X)\Psi _{+}^{(\sigma )}(x)=0,\hspace{0.5in}(D_\mu
^2-m^2-a\sigma X^{*})\Psi _{-}^{(\sigma )}(x)=0,
\end{equation}
where $a=eq/s.$ These equations are like the Klein-Gordon equation for
scalar particles but with complex ''effective'' masses: $m_{eff}^2=m^2+a
\sigma X,$ $\left( m_{eff}^2\right) ^{*}=m^2+a\sigma X^{*}.$ It is
sufficient to consider only one of equations (40). Let us consider the
solution of the equation

\begin{equation}
(D_\mu ^2-m_{eff}^2)\Psi ^{(\sigma )}(x)=0,\hspace{1.0in}(\Psi ^{(\sigma
)}(x)\equiv \Psi _{+}^{(\sigma )}(x))
\end{equation}
in the presence of the external constant uniform electromagnetic fields. The
general case is when two Lorentz invariants of the electromagnetic fields $
{\cal F}=\frac 14F_{\mu \nu }^2\neq 0,$ ${\cal G}=\frac 14F_{\mu \nu } 
\widetilde{F}_{\mu \nu }\neq 0$ ($\widetilde{F}_{\mu \nu }=\frac i2\epsilon
_{\mu \nu \alpha \beta }F_{\alpha \beta },$ $\epsilon _{\mu \nu \alpha \beta
}$ is the antisymmetric Levi-Civita tensor). Then there is a coordinate
system where the electric ${\bf E}$ and magnetic ${\bf H}$ fields are
parallel, i.e. ${\bf E\parallel H}.$ In this case the 4-vector potential is
given by
\begin{equation}
A\mu =\left( 0,x_1H,-x_0E,0\right)
\end{equation}
so that 3-vectors ${\bf E}={\bf n}E,$ ${\bf H}={\bf n}H$ directed along the
axes 3, where ${\bf n}=(0,0,1)$ is a unit vector. The four solutions of
equation (41) for the potential (42) with different asymptotic are given by
[21, 37] (see also [38])
\begin{equation}
_{\pm }^{\pm }\Psi _{p,n}^{(\sigma )}(x)=N\exp \left\{ i(p_2x_2+p_3x_3)- 
\frac{\eta ^2}2\right\} H_n(\eta )_{\pm }^{\pm }\psi ^{(\sigma )}(\tau )
\end{equation}
where $N$ is the normalization constant, $H_n(\eta )$ is the Hermite
polynomial, 
\[
\eta =\frac{eHx_1+p_2}{\sqrt{eH}},\hspace{0.5in}\nu =\frac{ik^2}{2eE}-\frac
12, \hspace{0.5in}\tau =\sqrt{eE}\left( x_0+\frac{p_0}{eE}\right) 
\]
and 
\[
_{+}^{}\psi ^{(\sigma )}(\tau )=D_\nu [-(1-i)\tau ]\hspace{1.0in}_{}^{-}\psi
^{(\sigma )}(\tau )=D_\nu [(1-i)\tau ] 
\]
\begin{equation}
_{}^{+}\psi ^{(\sigma )}(\tau )=D_{\nu ^{*}}[(1+i)\tau ]\hspace{1.0in}
_{-}^{}\psi ^{(\sigma )}(\tau )=D_{\nu ^{*}}[-(1+i)\tau ]
\end{equation}

Here $D_\nu (x)$ is the Weber-Hermite function (the parabolic-cylinder
function). The probability for pair production of particles with arbitrary
spins by the constant electromagnetic fields can be obtained through the
asymptotic of solutions (44) when the time $x_0\rightarrow \pm \infty .$ The
functions $_{+}^{+}\psi ^{(\sigma )}(\tau )$ have positive frequency at $
x_0\rightarrow \pm \infty $ and $_{-}^{-}\psi ^{(\sigma )}(\tau )$ -
negative frequency. The constant $k^2$ which enters the index $\nu $ of the
parabolic-cylinder functions (44) is given by [38]
\begin{equation}
k^2=m_{eff}^2+eH(2n+1),
\end{equation}
where $n=l+r,$ $l$ is the orbital quantum number, $r$ is the radial quantum
number and $n=0,1,2,...$ is the principal quantum number. It should be noted
that for scalar particles we have the equation $k^2=p_0^2-p_3^2$, where $p_0$
is the energy and $p_3$ is the third projection of the momentum of a scalar
particle. In our case of arbitrary spin particles, the parameter $m_{eff}^2$
is the complex value. Nevertheless all physical quantities in this case are
the real values. Solutions (43), (44) are characterized by three conserved
numbers: $k^2$ and the momentum projections $p_2$, $p_3.$ As shown in [21]
functions (43) are connected by the relations 
\[
_{+}^{}\Psi _{p,n}^{(\sigma )}(x)=c_{1n\sigma }^{}{}^{+}\Psi _{p,n}^{(\sigma
)}(x)+c_{2n\sigma }^{}{}^{-}\Psi _{p,n}^{(\sigma )}(x), 
\]
\[
_{}^{+}\Psi _{p,n}^{(\sigma )}(x)=c_{1n\sigma }^{*}{}_{+}\Psi
_{p,n}^{(\sigma )}(x)-c_{2n\sigma }^{}{}_{-}^{}\Psi _{p,n}^{(\sigma )}(x), 
\]
\begin{equation}
_{}^{-}\Psi _{p,n}^{(\sigma )}(x)=-c_{2n\sigma }^{*}{}_{+}\Psi
_{p,n}^{(\sigma )}(x)+c_{1n\sigma }^{}{}_{-}^{}\Psi _{p,n}^{(\sigma )}(x),
\end{equation}
\[
_{-}^{}\Psi _{p,n}^{(\sigma )}(x)=c_{2n\sigma }^{*}{}^{+}\Psi
_{p,n}^{(\sigma )}(x)+c_{1n\sigma }^{*}{}^{-}\Psi _{p,n}^{(\sigma )}(x), 
\]
where coefficients $c_{1n\sigma }^{},$ $c_{2n\sigma }^{}$ are given by 
\[
c_{2n\sigma }^{}=\exp \left[ -\frac \pi 2(\lambda +i)\right] \hspace{1.0in}
\lambda =\frac{m_{eff}^2+eH(2n+1)}{eE}, 
\]
\[
\mid c_{1n\sigma }^{}\mid ^2-\mid c_{2n\sigma }^{}\mid ^2=1\hspace{1.0in}
for~ even~ spins, 
\]
\begin{equation}
\mid c_{1n\sigma }^{}\mid ^2+\mid c_{2n\sigma }^{}\mid ^2=1\hspace{1.0in}
for~ odd~ spins.
\end{equation}

The values $c_{1n\sigma }^{},$ $c_{2n\sigma }^{}$ are connected with the
probability of the pair producing of arbitrary spin particles in the state
with the quantum number $n$ and the spin projection $\sigma $. The absolute
probability for a production of a pair in the state with quantum number $n$,
momentum $p$ and the spin projection $\sigma $ in all space and during all
time is
\begin{equation}
\mid c_{2n\sigma }\mid ^2=\exp \left\{ -\pi \left[ \frac{m^2}{eE}+\frac{
q\sigma H}{sE}+\frac HE(2n+1)\right] \right\} .
\end{equation}

The value (48) is also the probability of the annihilation of a pair with
quantum numbers $n,$ $p,$ $\sigma $ with the energy transfer to the external
electromagnetic fields. It is seen from (48) that for $H\gg E$ the pair of
particles are mainly created by the external fields in the state with $n=0,$ 
$\sigma =-s.$ This is the state with the smallest energy. So at $H\gg E$
there is a production of polarized beams of particles and antiparticles with
the spin projection $\sigma =-s.$ ( the s is the spin of particles). The
average number of produced pairs of particles from a vacuum is
\begin{equation}
\overline{N}=\int \sum_{n,\sigma }\mid c_{2n\sigma }\mid ^2dp_2dp_3\frac{L^2 
}{(2\pi )^2}
\end{equation}
because $(2\pi )^{-2}dp_2dp_3L^2$ is the density of final states, where the $
L$ is the cut-off along the coordinates, so the $L^3$ is the normalization
volume. The variables $\eta ,$ $\tau $ define the region of forming the
process which is described by solutions (43) with the coordinates of the
centre of this region $x_0=-p_3/eE,$ $x_1=-p_2/eH.$ Therefore instead of the
integration over $p_2$ and $p_3$ in (49) it is possible to use the
substitution [21]
\begin{equation}
\int dp_2\rightarrow eHL,\hspace{1.0in}\int dp_3\rightarrow eET
\end{equation}
with the time of observation $T$.

Calculating the sum in (49) over the principal quantum number $n$, with the
help of (48), (50) we obtain the probability of the pair production per unit
volume and per unit time
\begin{equation}
I(E,H)=\frac{\overline{N}}{VT}=\frac{e^2EH}{8\pi ^2}\frac{\exp \left[ -\pi
m^2/(eE)\right] }{\sinh \left( \pi H/E\right) }\sum_\sigma \exp (-\sigma b),
\end{equation}
where $b=\pi qH/(sE),$ $V=L^3.$ Now we calculate the sum over the spin
projection $\sigma $ in (51) for odd and even spins:

1) even spins 
\[
\sum_\sigma \exp (-\sigma b)=S_1+S_{2,}\hspace{0.3in}
S_1=e^0+e^{-b}+...+e^{-sb}=\frac{e^{-b(s+1)}-1}{e^{-b}-1}, 
\]

\begin{equation}
S_2=e^b+e^{2b}+...+e^{sb}=\frac{e^b\left( e^{bs}-1\right) }{e^b-1},S_1+S_2= 
\frac{\cosh (sb)-\cosh \left[ (s+1)b\right] }{1-\cosh b}
\end{equation}

2) odd spins

\[
\sum_\sigma \exp (-\sigma b)=S_1^{^{\prime }}+S_2^{^{\prime }},\hspace{0.2in}
S_1^{^{\prime }}=e^{-b/2}+e^{-3b/2}+...+e^{-sb}=\frac{e^{-b(s+1)}-e^{-b/2}}{
e^{-b}-1}, 
\]

\begin{equation}
S_2^{^{\prime }}=e^{b/2}+e^{3b/2}+...+e^{sb}=\frac{e^{b(s+1)}-e^{b/2}}{e^b-1}
,S_1^{^{\prime }}+S_2^{^{\prime }}=\frac{\cosh (sb)-\cosh \left[
(s+1)b\right] }{1-\cosh b}
\end{equation}

So we get the same final expressions for even and odd spins. Using the
relationship
\begin{equation}
\frac{\cosh (sb)-\cosh \left[ (s+1)b\right] }{1-\cosh b}=\cosh (sb)+\sinh
(sb)\coth \frac b2=\frac{\sinh \left[ b(s+1/2)\right] }{\sinh (b/2)}
\end{equation}
and equations (51), (52) we arrive at the pair-production probability
\begin{equation}
I(E,H)=\frac{\overline{N}}{VT}=\frac{e^2EH}{8\pi ^2}\frac{\exp \left[ -\pi
m^2/(eE)\right] }{\sinh \left( \pi H/E\right) }\frac{\sinh \left[ (2s+1)q\pi
H/(2sE)\right] }{\sinh \left[ q\pi H/(2sE)\right] }.
\end{equation}

Expression (55) coincides with those derived by [23] using the
quasiclassical approach. So the $I(E,H)$ is the intensity of the creation of
pairs of arbitrary spin particles which possess the magnetic moment $\mu
=eq/(2m)$ and gyromagnetic ratio $g=q/s.$

In [23] there is a discussion of physical consequences which follow from
equation (55). In particular, there is a pair production in pure magnetic
field if $q=gs>1$ [23]. It is interesting that the exact formula derived
here from quantum field theory which is valid for arbitrary fields $E,$ $H,$
coincides with the asymptotic expression obtained by [23] for $E,H\ll m^2/e.$

To get the imaginary part of the density of the Lagrangian we use the
relationship~[21]

\begin{equation}
VTIm{\cal L}=\frac 12\int \sum_{n,\sigma }\ln \mid c_{1n\sigma }\mid
^2dp_2dp_3\frac{L^2}{(2\pi )^2}.
\end{equation}
With the help of (47), (50) we arrive at (see also [23])
\begin{equation}
Im{\cal L}=\frac{e^2EH}{16\pi ^2}\sum_{n=1}^\infty \frac{\beta _n}n\exp
\left( -\frac{\pi m^2n}{eE}\right) \frac{\sinh \left[ n(2s+1)q\pi
H/(2sE)\right] }{\sinh \left( n\pi H/E\right) \sinh \left[ nq\pi
H/(2sE)\right] },
\end{equation}
where 
\[
\beta _n= \left\{
\begin{array}{c}
(-1)^{n-1}\hspace{0.75in}for~bosons \\ 
\hspace{3em}1\hspace{0.8in}for~fermions.
\end{array}\right.
\]

The different expressions for bosons and fermions occur due to different
statistics and relations (47). The first term ($n=1$) in (57) coincides with
the probability of the pair production per unit volume per unit time divided
by 2 [22] (see discussion in [23]).

\section{Vacuum Polarization of Arbitrary Spin Particles}

Now we calculate the nonlinear corrections to the Lagrangian of a constant
uniform electromagnetic field interacting with the vacuum of arbitrary spin
particles with the gyromagnetic ratio $g$. For the case of spins $0$, $1/2$
and $1$ (for $g=2$) such problem was solved by authors [39, 40, 22, 24]. The
nonlinear corrections to Lagrangian of the electromagnetic field describe
the effect of scattering of light by light. We consider one loop corrections
corresponding to arbitrary spin particles to the Maxwell Lagrangian. For
this purpose to take into account the vacuum polarization, it is convenient
to explore the Schwinger method [22]. Applying this approach to the
arbitrary spin particles described by equation (1) we arrive at the
effective Lagrangian of constant uniform electromagnetic fields
\begin{equation}
{\cal L}_1=-\frac 1{32\pi ^2}\int_0^\infty d\tau \tau ^{-3}\exp \left(
-m^2\tau -l(\tau )\right) tr\exp \left( \frac{eq}{2s}\Sigma _{\mu \nu
}F_{\mu \nu }\tau \right) ,
\end{equation}
where
\begin{equation}
\Sigma _{\mu \nu }=\Sigma _{\mu \nu }^{(+)}\oplus \Sigma _{\mu \nu }^{(-)}, 
\hspace{0.5in}l(\tau )=\frac 12tr\ln \left[ \left( eF\tau \right) ^{-1}\sin
(eF\tau )\right]
\end{equation}
and $F_{\mu \nu }$ is a constant tensor of electromagnetic fields. The
formal different of (60) from the case of spin $1/2$ particles is in the
substitution $\sigma _{\mu \nu }\rightarrow (q/s)\Sigma _{\mu \nu },$ where $
\sigma _{\mu \nu }=(i/2)\left[ \gamma _\mu ,\gamma _\nu \right] ,$ $\gamma
_\mu $ are the Dirac matrices. The problem is to calculate the trace of
matrices entering the exponential factor in (58). Using relations (37)-(39),
(52)-(54) we find
\begin{equation}
tr\exp \left( \frac{eq}{2s}\Sigma _{\mu \nu }F_{\mu \nu }\tau \right)
=2Re\left[ \cosh (eqX\tau )+\sinh (eqX\tau )\coth \left( \frac{eqX\tau }{2s}
\right) \right] .
\end{equation}

Inserting (61) into (60) and adding the constant which is necessary to
cancel ${\cal L}_1$ when electromagnetic fields are turned off (see [22]) we
arrive at 
\[
{\cal L}_1=-\frac 1{8\pi ^2}\int_0^\infty d\tau \tau ^{-3}\exp \left(
-m^2\tau \right) \times 
\]
\begin{equation}
\times \left[ (e\tau )^2{\cal G}\frac{Re\left[ \cosh (eqX\tau )+\sinh
(eqX\tau )\coth \left( eqX\tau /(2s)\right) \right] }{2Im\cosh (eX\tau )}- 
\frac{2s+1}2\right] ,
\end{equation}
where ${\cal G}={\bf EH}$. At $q=1$ and $s=1/2$ Lagrangian (61) coincides
with the Schwinger one [22]. Expression (61) is the correction to the
Maxwell Lagrangian with taking into account the vacuum polarization of
arbitrary spin particles which possesses the magnetic moment $\mu =eq/(2m)$
and gyromagnetic ratio $g=q/s.$ Adding (61) to the Lagrangian of the free
electromagnetic fields 
\[
{\cal L}_0=-{\cal F=}\frac 12\left( {\bf E}^2-{\bf H}^2\right) 
\]
and introducing the divergent constant for weak fields, we get the
expression for the total Maxwell Lagrangian 
\[
{\cal L}_M={\cal L}_0+{\cal L}_1=-Z{\cal F}-\frac 1{8\pi ^2}\int_0^\infty
d\tau \tau ^{-3}\exp \left( -m^2\tau \right) \times 
\]
\begin{equation}
\times \left[ (e\tau )^2{\cal G}\frac{Re\left[ \cosh (eqX\tau )+\sinh
(eqX\tau )\coth \left( eqX\tau /(2s)\right) \right] }{2Im\cosh (eX\tau )}- 
\frac{2s+1}2-4\beta (e\tau )^2{\cal F}\right] ,
\end{equation}
where
\begin{equation}
Z=1+\frac{e^2\beta }{2\pi ^2}\int_0^\infty d\tau \tau ^{-1}\exp \left(
-m^2\tau \right) ,\hspace{0.3in}\beta =\frac{q^2(2s^2+3s+1)-s(2s+1)}{24s}.
\end{equation}

Following the Schwinger procedure, the renormalization of electromagnetic
fields ${\cal F}\rightarrow Z{\cal F}$ and charge $e\rightarrow Z^{-1/2}e$
are used. After expanding (63) in the small electric $E$ and magnetic $H$
fields we arrive at the Lagrangian of constant uniform electromagnetic
fields (in rational units)
\begin{equation}
{\cal L}_M=\frac 12\left( {\bf E}^2-{\bf H}^2\right) + \frac{2\alpha ^2}{
45m^4}\left[ \left( {\bf E}^2-{\bf H}^2\right) ^2(15\beta -\gamma )+({\bf EH})^2\left( 4\gamma +\frac{2s+1}2\right) \right] +...
\end{equation}
where $\alpha =e^2/(4\pi )$ and
\begin{equation}
\gamma =\frac{\left[ q^4\left( 6s^4+15s^3+10s^2-1\right) -3s^3\left(
2s+1\right) \right] }{16s^3}.
\end{equation}

It is easy to check that as a particular case at $s=1/2,$ $q=1$ (which
corresponds to the Dirac theory), (65) coincides with the well-known
Schwinger Lagrangian [22]. At $s=1$ and $q=1$ expression (64) is different
from one received in [24]. It is because the considered theory of particles
at $s=1$ is not equivalent to the Proca theory. There is the doubling of
spin states here. Effective Lagrangian (65) is the Heisenberg-Euler type
which has been found for the case of polarization vacuum of particles with
arbitrary spin and magnetic moment. Here we took into account virtual
arbitrary spin particles but not virtual photons. It is because at small
energies of the external fields the radiative corrections are small
quantities. It is not difficult to find the asymptotic of (62) for over
critical fields $eE/m^2\rightarrow \infty $ and $eH/m^2\rightarrow \infty $
. It should be noted, however, that for strong electromagnetic fields
anomalous magnetic moment of electrons depend on external fields [41, 42]
and so do for arbitrary spin particles. Therefore to make the correct limit
it is necessary to take into account this dependence [26].

It is possible to receive also the imaginary part of Lagrangian (57) from
(62) using the residue theorem taking into account of poles of expression
(62) and passing above them [22].

\section{Discussion of the Results}

The theory of particles with arbitrary spins and magnetic moment based on
equation~(1) and the corresponding Lagrangian allow to find density matrices
(34), (35), pair-production probability (55) and effective Lagrangian for
electromagnetic fields (65) also taking into account the polarization of
vacuum. It is convenient to use matrices-diads (34), (35) for different
electrodynamics calculations with the presence of particles with arbitrary
spins. The exact formula for the intensity of pair production of arbitrary
spin particles coincides with the expression obtained by [23] using the
quasiclassical method of ''imaginary time'' which is valid only for $E,$ $
H\ll m^2/e,$ i.e. for weak fields. From this follows that the analysis made
in [23] is valid for arbitrary electromagnetic fields and is grounded by the
relativistic quantum field theory. In particular, there is a pair production
by a pure magnetic field $gs>1$ [23] and in the presence of the magnetic
field the probability decreases for scalar particles and increases for
higher spin particles. As all divergences and the renormalizability are
contained in $Re{\cal L}$ (62) but not in $Im{\cal L}$, the pair production
probability does not depend on scheme of renormalizability. The cases of
scalar and spinor (with $g=2$) particles are reliable for calculations of
vacuum polarization corrections due to their theories are renormalizable.
The general formula (62) obtained here presents interest for the further
development of the field theory particles with higher spins (see discussion
in [23], [24]). Expression (62) is a reasonable result for arbitrary values
of $s$ and $q$ because for particular case of scalar and spinor particles we
arrive at known result. So we have here a reasonable and non-contradictory
description of the nonlinear effects that arise in this interaction.

\end{document}